\documentclass[conference]{IEEEtran}
\usepackage[export]{adjustbox} 
\usepackage{ragged2e}
\usepackage[utf8]{inputenc}
\usepackage{fourier} 
\usepackage{array}
\usepackage{makecell}
\usepackage{multirow}
\usepackage{siunitx}
\usepackage{multicol}
\usepackage{amsmath}

\IEEEoverridecommandlockouts
\usepackage{cite}
\usepackage{amsmath,amssymb,amsfonts}
\usepackage{algorithmic}
\usepackage{graphicx}
\usepackage{textcomp}
\usepackage{xcolor}
\def\BibTeX{{\rm B\kern-.05em{\sc i\kern-.025em b}\kern-.08em
    T\kern-.1667em\lower.7ex\hbox{E}\kern-.125emX}}
\begin{document}

\title{Optimizing Electric Vehicle Efficiency with Real-Time Telemetry using Machine Learning\\

}

\author{
\IEEEauthorblockN{Aryaman Rao}
\IEEEauthorblockA{\textit{dept. of Electrical Engineering} \\
\textit{Delhi Technological University  }\\
Delhi, India \\
aryaman26601@gmail.com}
\and
\IEEEauthorblockN{Harshit Gupta}
\IEEEauthorblockA{\textit{dept. of Mechanical Engineering  } \\
\textit{Delhi Technological University  }\\
Delhi, India \\
hgh1602@gmail.com}
\and
\IEEEauthorblockN{Parth Singh}
\IEEEauthorblockA{\textit{dept. of Mechanical Engineering  } \\
\textit{Delhi Technological University  }\\
Delhi, India \\
parth55singh@gmail.com}
\and
\IEEEauthorblockN{Shivam Mittal}
\IEEEauthorblockA{\textit{dept. of Applied Mathematics } \\
\textit{Delhi Technological University }\\
Delhi, India \\
shivammittal2124@gmail.com}
\and
\IEEEauthorblockN{Utkarsh Singh}
\IEEEauthorblockA{\textit{dept. of Mechanical Engineering } \\
\textit{Delhi Technological University }\\
Delhi, India \\
utkarsh.an23o5@gmail.com}
\and
\IEEEauthorblockN{Dinesh Kumar Vishwakarma}
\IEEEauthorblockA{\textit{dept. of Information Technology} \\
\textit{Delhi Technological University}\\
Delhi, India \\
dinesh@dtu.ac.in}
}


\maketitle
\begin{abstract}
In the contemporary world with degrading natural resources, the urgency of energy efficiency has become imperative due to the conservation and environmental safeguarding. Therefore, it's crucial to look for advanced technology to minimize energy consumption. This research focuses on the optimization of battery-electric city style vehicles through the use of a real-time in-car telemetry system that communicates between components through the robust Controller Area Network (CAN) protocol. By harnessing real-time data from various sensors embedded within vehicles, our driving assistance system provides the driver with visual and haptic actionable feedback that guides the driver on using the optimum driving style to minimize power consumed by the vehicle. To develop the pace feedback mechanism for the driver, real-time data is collected through a Shell Eco Marathon Urban Concept vehicle platform and after pre-processing, it is analyzed using the novel machine learning algorithm TEMSL, that outperforms the existing baseline approaches across various performance metrics. This innovative method after numerous experimentation has proven effective in enhancing energy efficiency, guiding the driver along the track, and reducing human errors. The driving-assistance system offers a range of utilities, from cost savings and extended vehicle lifespan to significant contributions to environmental conservation and sustainable driving practices.\\
\end{abstract}

\vspace{-5pt}
\section{Introduction}

In recent times, the significance of energy efficiency has grown monumentally, driven by the urgent necessity to address waste management and mitigate pollution for a sustainable environment. This alarming concern extends to various aspects of daily life, including the transportation industry \cite{b1}. The pollution emitted by the increasing number of gasoline vehicles present on the road is nearly unavoidable and the amount of greenhouse gases released by running vehicles standstill at a traffic light is an understatement to the impact they have on the environment\cite{b2}. Besides, how the car is controlled, particularly the way it accelerates, brakes, and maneuvers can have a significant impact on its overall driving efficiency. Even a minor malfunction in any component of the vehicle can cost its efficiency heavily. While significant research has been conducted on increasing vehicle efficiency\cite{b3}, there is still scope for further improvement with the advent of artificial intelligence in the automotive industry.

With so many variables at play when driving, it's challenging for the driver to always operate under optimal conditions. That's where a real-time telemetry system comes into play. It can assist the driver by providing important insights about the behaviour and performance changes of the car. Our system utilizes different sensors to gather raw data during various runs and after thorough data analysis, we gain an understanding of the car's behavior, its component interactions, and especially energy consumption.

The Shell Eco-Marathon is an annual global student competition where student teams design and fabricate their vehicles to compete against each other to achieve the highest possible energy efficiency. The Urban Concept class represents a replica of modern-day passenger vehicles and is one of the different categories of vehicles eligible for the competition. , which are predominantly designed to attain exceptional energy-efficiency. As shown in the Fig. \ref{Test}, our battery-electric Urban Concept vehicle participated in the International Shell Eco-Marathon Asia-Pacific Competition held on October 2022 at Lombok, Indonesia.


Through this paper, we delve into the progression and working of the telemetry system with emphasis on the novelty and research characteristics of the system employed in a battery-powered Urban Concept vehicle. Stress is given on how this evidence-based methodology, fortified by machine learning (ML) integration, can lead to a more efficient racetrack experience. This research work is part of a larger effort to save energy and make urban travel better for a more promising future. Our concept received the 2\textsuperscript{nd} prize in the Data and Telemetry Award, sponsored by Shell and Schmid Elektronik, acknowledging the innovative contribution and advocating the use of data to achieve sustainable mobility.


The subsequent sections are organised as follows. Section II involves literature review and finding out existing research gaps that our work suffices while in Section III, we explain the set up and methodology of our experiment which includes details about hardware, driving strategies, and the onboard systems. In Section IV, we provide information about the various data processing techniques and machine learning models utilized and in Section V, we have present the simulation outcomes and key findings, followed by conclusions in Section VI.

\section{Literature Review}
With electric vehicles (EVs) taking centre stage in the consumer market, significant advancements have been made in enhancing their efficiency through exterior remodelling or introducing semi-autonomous (upto Level 3) driving capabilities \cite{Parekh}. These strategies utilize diverse vehicle data points in order to achieve any efficacious results \cite{Vasudevan}. With the current market trend being driver engagement minimization while driving, little research is being done on learning based predictive control strategies.\\
In the context of forward-looking driving assistance for energy efficiency, systems that provide drivers with insights about speed limits have been developed. This early awareness helps drivers avoid undesired braking losses and offers guidance for energy-efficient driving \cite{b4, b5}. However, a common challenge with most modern methods is that they rely on internet access for real-time updates or access to traffic conditions and might not work satisfactorily in areas with poor internet connectivity. Our telemetry system builds upon this by using CAN protocol for communication instead of popular wireless protocols such as Bluetooth, WiFi and LTE \cite{Triwiyatno,b8,b12}. An interesting area to explore here involves developing driver assistance systems that promote efficient driving. These systems use techniques like shared control, adjusting air-to-fuel ratios, and adapting speed based on the type of terrain \cite{Ivarsson, Santana}. Multiple studies have even tried to classify different driving patterns and environments \cite{Meseguer}. Yet, drivers often lack actionable feedback to help them improve driving efficiency.\\
With the recent advances in Artificial Intelligence over the past few decades, many researchers have been eager to incorporate them into various fields. Some researchers have dived into using neural networks to improve vehicle efficiency, as discussed in \cite{Mosabbir}. Several studies \cite{Pan22} have been conducted on extracting valuable information using feature engineering on time series problems. Ensemble modelling has become quite popular in predicting patterns in telemetry or time-variant data, as demonstrated in \cite{Ravikumar20}.The potential of integrating onboard sensors and real-time data processing is intriguing for the realm of vehicle efficiency. These sensors are vital in delivering critical insights into factors such as the road surface characteristics, traffic density, and gradient variations. These critical insights facilitate the implementation of precise and context-aware energy management strategies \cite{Kohlhaas11, Wickramanayake}, ultimately contributing to the improvement in operational efficiency of the vehicle. Harnessing this data empowers vehicles to dynamically fine-tune power distribution, optimize regenerative braking mechanisms, and provide optimal route suggestions. Furthermore, the development of vehicle-to-vehicle (V2V) and vehicle-to-infrastructure (V2I) communication systems \cite{Rayamajhi16} presents a promising avenue for further vehicle efficiency optimization. These systems facilitate seamless data exchange on aspects such as speed, location, and driver intent, enabling the deployment of collaborative strategies aimed at minimizing energy-depleting actions such as abrupt lane changes and sudden acceleration. The pace feedback mechanism proposed by us expands upon the existing literature by providing an accessible haptic and visual feedback interface based on robust CAN communication enabling the driver to easily maximize vehicle efficiency by altering their driving style.

\begin{figure}[t]
    \centering
    \includegraphics[scale = 0.22]{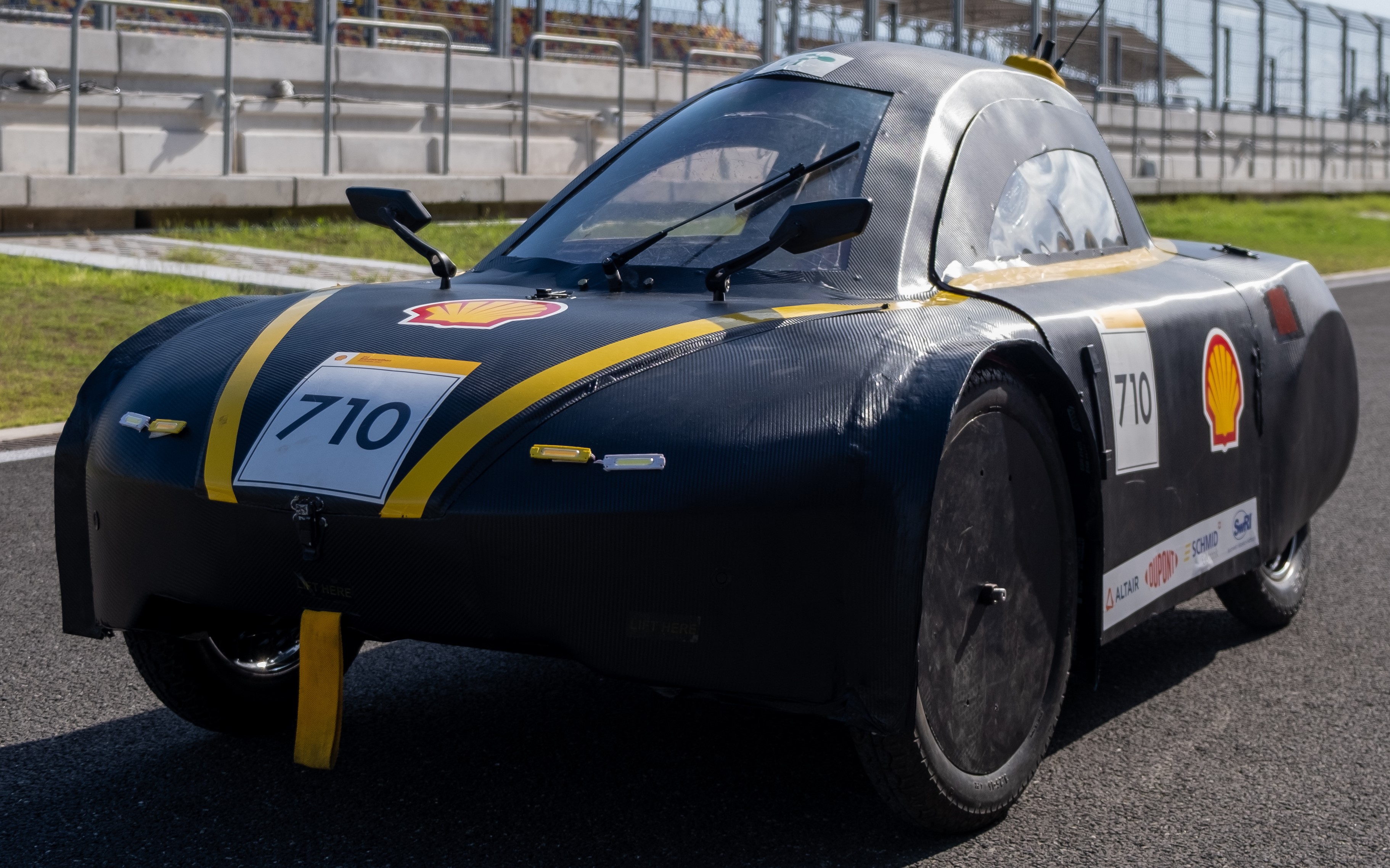}
    \caption{Test Vehicle}
    \label{Test}
\end{figure}

\section{Data Acquisition}
The real-time data acquisition setup for research into the optimum driving strategy consists of various elements such as sensors ranging from accelerometers to joule-meters and a robust network. Various key aspects such as the layout of the track, the driving methods employed, and the inclusion of special equipment for real-time telemetry on the vehicle are discussed. An insight into the methods employed to acquire data is necessary for the reliability and dependability of the mechanism.


\subsection{Test Track}\label{AA}
To test the working of the telemetry system, we had to
perform a number of test laps in the Urban Concept vehicle mimicking the actual track to be encountered in Shell Eco Marathon.
The test track is shown in the Fig.\ref{Test} highlighted in blue. The
route was created inside the Delhi Technological University
campus with two objectives in mind.\\
\vspace{-7pt}
\begin{figure}
    \centering
    \includegraphics[scale = 0.25]{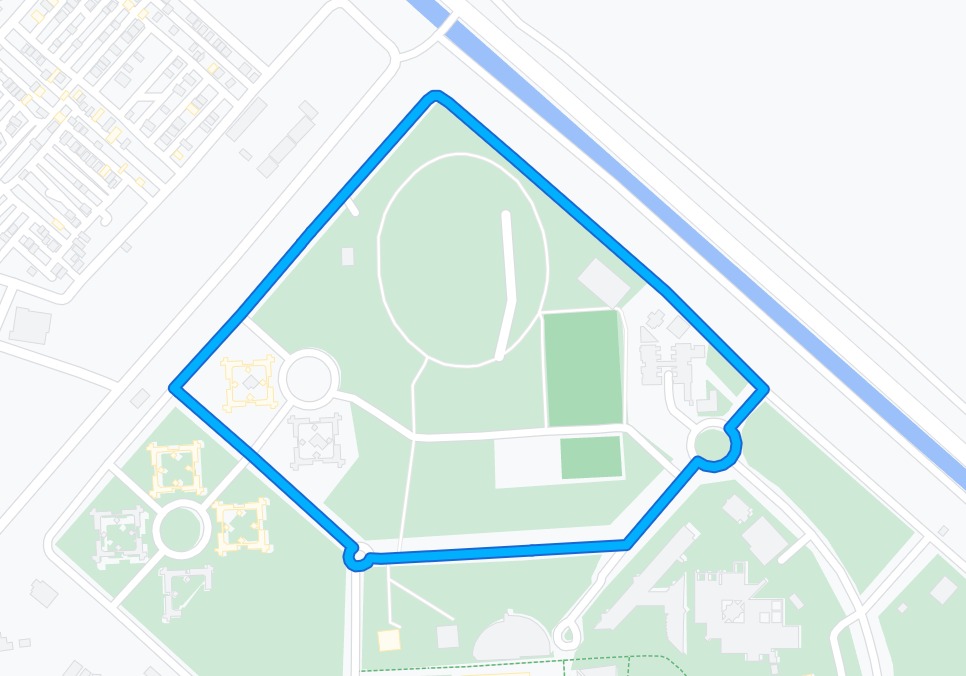}
    \caption{Test Track}
    \label{Test}
\end{figure}

\subsubsection{Track Length}
The length of the track is chosen as close as possible to the
Shell Eco-Marathon tracks, to provide data that will be
helpful on track. The average length of a Shell Eco-Marathon track is 1.4 km therefore, the route chosen to be covered was about 1.5 km long.\\
\vspace{-7pt}
\subsubsection{Number of Corners}
This is a critical parameter to understand and devise an optimum pattern with the objective of minimizing energy consumption. Our test track had 8 corners over which the vehicle was tested for 24 laps in slightly less than 2 hours wherein driving strategies were changed after every 4 laps. Data was collected every 100 ms for the entire run, amounting to more than 10,000 data points which were split in the ratio of 7:3, comprising of 70\% data for training and 30\% data for testing.
\subsection{Driving Techniques}
Driving behaviors and proper situation-dependent techniques can save energy consumption in a dynamic manner as it affects the rate of energy consumption and gas emission as much as the manufacturing technologies involved. These habits have significant effects on reducing energy consumption in EVs.\cite{Younes22}\\

\begin{table}[b!]
\caption{Acceleration Variation Method}
  \begin{center}
    \scalebox{0.85}{
    \begin{tabular}{ | c | c | c | c |}
      \hline
      \thead{Driving Style} & \thead{Extra Power Required \\ (As compared to Gentle)} & \thead{Energy Consumed \\ (As compared to Gentle)} \\
      \hline
      Aggressive &  \makecell{46.6 \%}  & 41.5 \%  \\
      \hline
      Mild &  \makecell{30.0 \%} & 15.7 \%  \\
      \hline
    \end{tabular}
    }
  \end{center}
\label{tb1}
\end{table}
\begin{table}[h]
\caption{Burn and Coast Method}
  \begin{center}
    \begin{tabular}{ | c | c |}
      \hline
      \thead{Driving Technique} & \thead{Efficiency Increase} \\
      \hline
      Full burn and coast &  \makecell{15 \%}  \\
      \hline
      Mild Burn and coast &  \makecell{8.5 \%}  \\
      \hline
    \end{tabular}
  \end{center}
\label{tb2}
\end{table}

\subsubsection{Acceleration Variation Method (Table \ref{tb1})}
In EVs, different driving styles can have great impact on the vehicle’s battery life, efficiency, and the SOC gradient. Before testing the vehicle some assumptions were made: there should be minimal traffic on the track and acceleration and jerk should only be done on the straight line paths, not around the corners. The different acceleration patterns that were tested are as follows:

\begin{itemize}
    \item {Aggressive: }
    Acceleration at a rapid rate
    \item {Gentle: }
    Acceleration at an optimal rate
    \item {Mild: }
    Acceleration at a slow rate
\end{itemize}

Several test runs were conducted for different patterns to plot a graph between acceleration and the power required from battery. \\

\begin{figure}[b]
    \centering
    \includegraphics[scale=0.22]{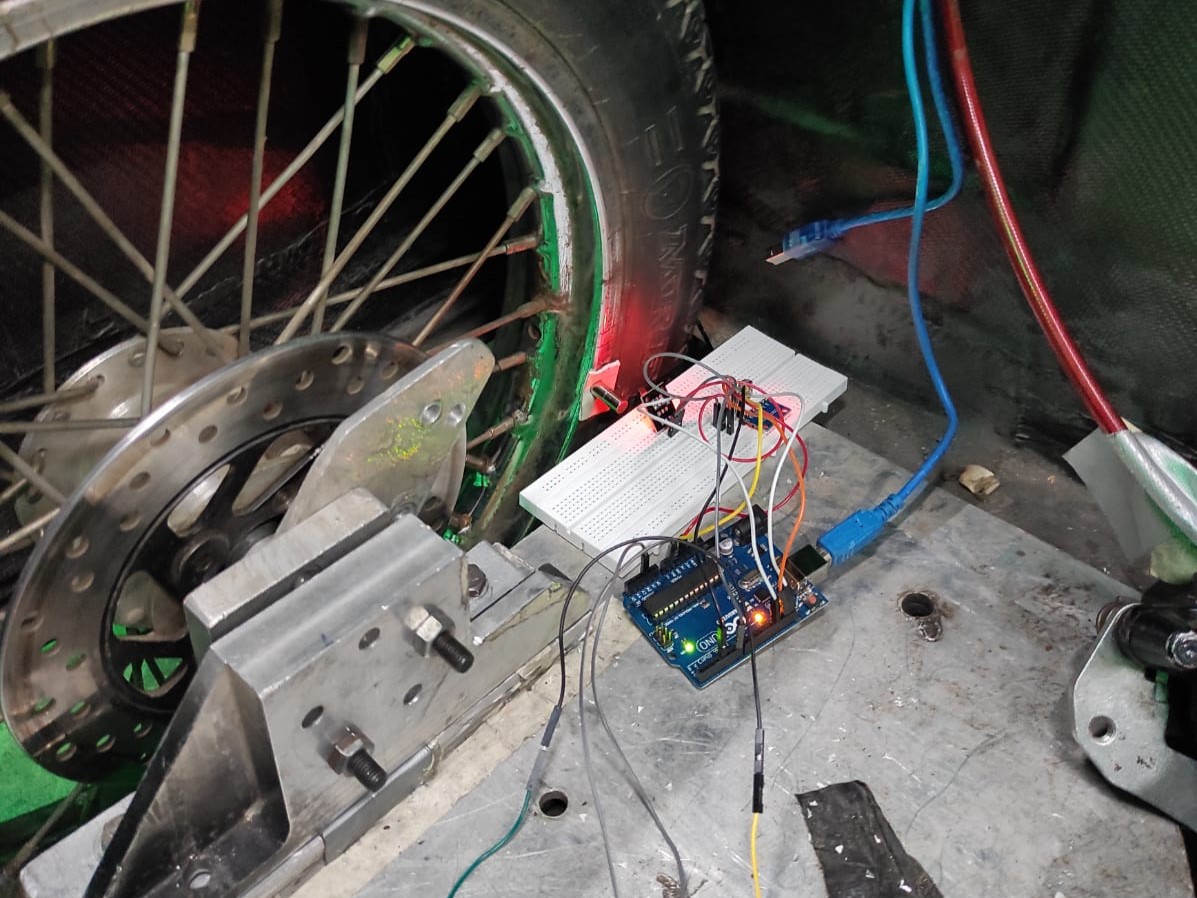}
    \caption{Raspberry Pi Setup in our Test Vehicle}
    \label{setup}
\end{figure}
\begin{figure}
    \centering
    \includegraphics[scale=0.2]{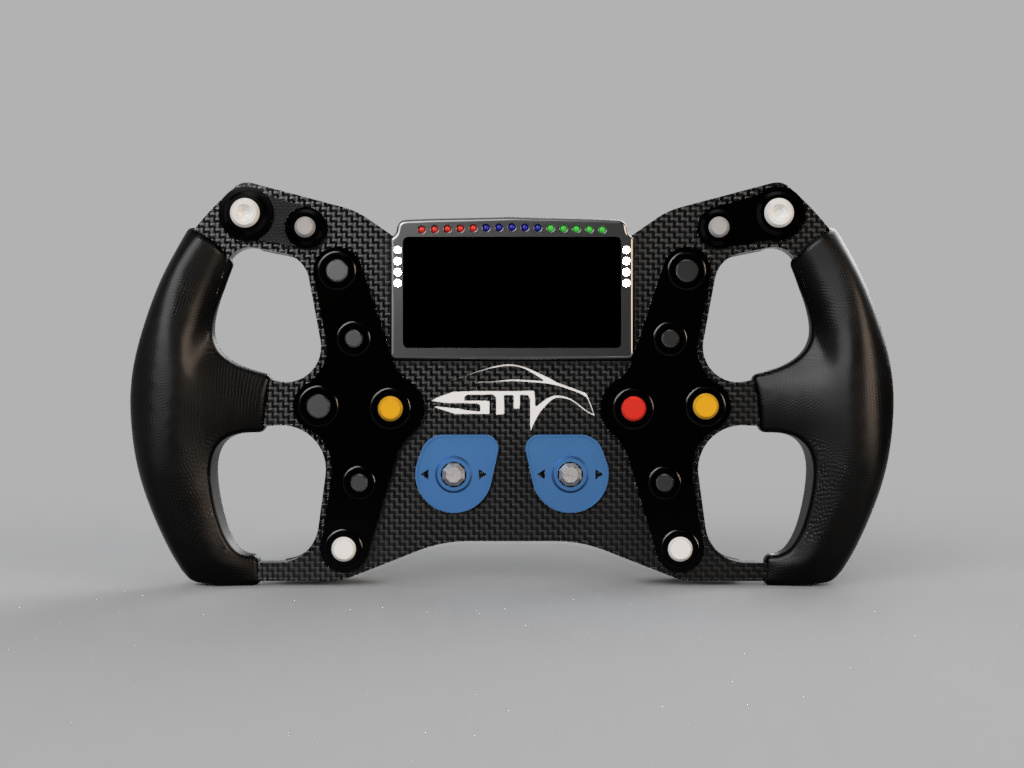}
    \caption{Working of the LEDs Pace Feedback on the Steering Wheel}
    \label{steering wheel}
\end{figure}


\subsubsection{Burn and Coast Method: (Table \ref{tb2}) }
It is a very popular driving method wherein the vehicle is accelerated to a particular speed, and the car is allowed to glide by switching off the engine till the vehicle speed drops to a set minimum velocity from where the motor is started again.
\begin{figure}[h!]
    \centering
    \includegraphics[scale = 0.35]{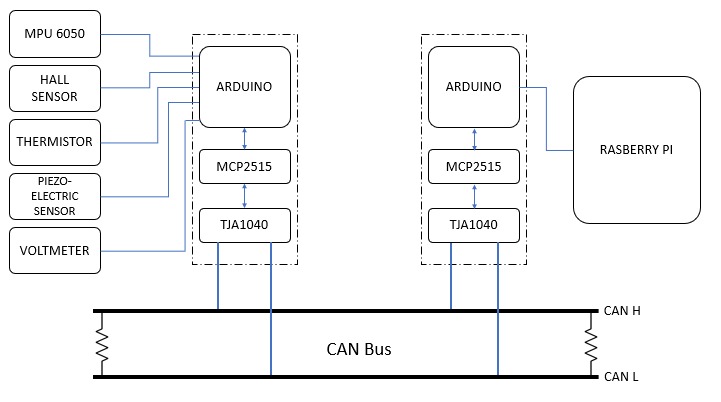}
    \caption{Block diagram depicting sensor-to-Raspberry Pi communication via CAN bus protocol, intermediated by Arduino for seamless data exchange}
    \label{block}
\end{figure}
Since our vehicle is battery-electric, we do not switch off the motor and we ensure to maintain a constant average velocity on straight paths as much as possible. The frequent switching on-and-off of the motor might induce high currents in the motor, consequently affecting it's efficiency, damaging it or harming the driver and the vehicle. Two techniques of burn and coast method have been investigated in the research study namely “Full” burn and coast and “mild” burn and coast. \cite{Jeongwoo}





\subsection{Onboard Telemetry System}
Sensors used to collect various data points for understanding the effect of driving techniques on the energy consumption in real-time are mentioned in Table \ref{Sensors} which are connected to their respective Arduino Uno modules. The nucleus of the entire telemetry system is a Raspberry Pi 4-4GB RAM board connected to the steering wheel. A 5-inch touch screen is also attached to the steering wheel to allow the driver to interact with the interface. Sensors can be connected directly using the I2C communication to the Raspberry Pi, or through the CAN Protocol as shown in figure \ref{block}. Sensors act as the clients and the Pi board acts as the centralized server, hence forming the Star Topology network. The CAN protocol is used to provide a robust communication between a controller and transceiver. MCP2515 acts as the CAN controller capable of transmitting and receiving data at high speeds (1 Mbps) while the TJA1040 creates an interface between the CAN controller and the bus. The combined data is sent over the CAN bus to the Raspberry Pi, where it is separated into individual data points. The steering wheel (Fig. \ref{steering wheel}) has been programmed to provide both visual and haptic feedback to the driver, using LEDs and two vibration motors under the finger rests; this system has been dubbed as the "Pace Feedback Mechanism". Slow velocity and acceleration are represented by RED lights with low intensity-low frequency haptics, high velocity and acceleration are indicated by GREEN lights with high intensity-high frequency haptics, and the optimal pace is denoted by BLUE with no haptic feedback.
\begin{center}
\begin{table}[hb]
\centering
\caption{Onboard Sensors for Data Acquisition}
\scalebox{1}{
\begin{tabular}{|c|c|c|}
    \hline
    \thead{Sr No.} & \thead{Sensors} & \thead{Functionality} \\ \hline
    1 & \makecell{MPU 6050 \cite{b14}} & Accelerometer \& Gyroscope  \\ 
    \hline
    2 & \makecell{Hall Sensor \cite{b15}} &  Speedometer \& Tachometer \\ 
    \hline
    3 & \makecell{NEO-6m Sensor \cite{b16}} & GPS receiver \\ 
    \hline
    4 & \makecell{Piezoelectric Sensor \cite{b17}} & Pressure\\ 
    \hline
    5 & \makecell{Motor-Controller} & Current \& Voltage from Battery.\\ 
    \hline
    6 & \makecell{Joulemeter} & Energy Consumed \\
    \hline
    7 & \makecell{NTC Thermistor \cite{b18}} & Temperature  \\ 
    \hline
\end{tabular}
}
\label{Sensors}
\end{table}
\end{center}

\section{Proposed Methodology \& Evaluation} 

\subsection{Data Mining}
Data mining research has made significant strides in temporal data analysis encompassing methods such as data representation, dimensionality reduction, and segmentation. In this study, we have integrated a comprehensive data mining framework, as explicated in the work of \cite{Liao12}. This framework was instrumental in exploiting the power of data mining, that leveraged various ML techniques, as cited by\cite{Griffiths, Constantinescu}, to extract the crucial information and insights from raw datasets. In accordance with the observations made by I. K. Nti \cite{Adekoya}, the omnipotence of data emphasizes the necessity for preprocessing methodologies that can adeptly harness information and knowledge. In our research, we have systematically partitioned this section into three integral components: Data Preprocessing, Modelling Framework, and Model Evaluation.
\begin{equation}
    \chi^2 = \sum \frac{(O - E)^2}{E}
\label{chi}
\end{equation}

Data preprocessing is a fundamental step in data analysis, essential for elevating the quality and reliability of predictive models. This process covers attributes like outlier rectification, management of missing data, and the normalization of features to ensure compatibility. Its importance is evident in real-world applications, where data often suffers from contamination by noise and intricate patterns. Through the application of such techniques, it can curtail the training complexity and significantly augment the efficacy of predictive analysis of telemetry data. A prevailing issue in recent ML models has been that of high dimensionality. To combat this drawback in large datasets, we apply feature extraction techniques to transform the data into a lower dimensional vector, making the model more efficient.  We employ Principal Component Analysis (PCA) \cite{Rosipal}, which is an unsupervised ML algorithm, that reduces dimensionality and improves performance for our experimentation.

\begin{figure}[hb]
    \centering
    \includegraphics[scale = 0.31]{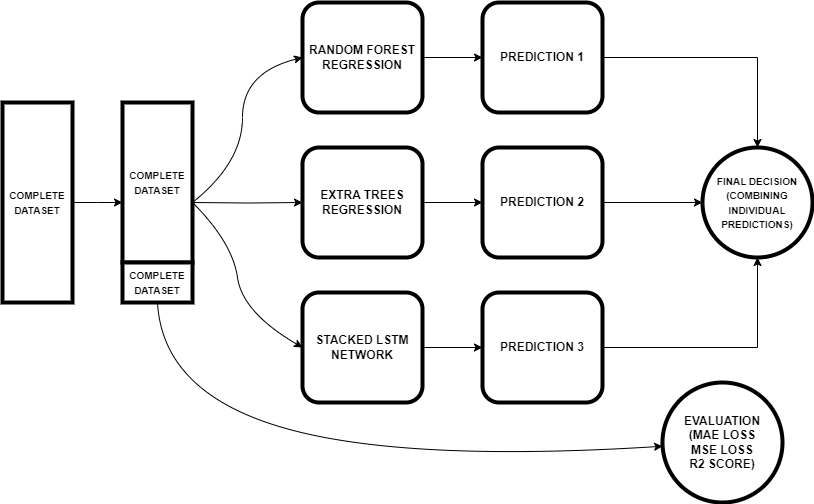}
    \caption{Model Architecture}
    \label{flowchart}
\end{figure}

Feature selection also plays a crucial role in enhancing the performance of multitude of ML models, assisting extracting relevant information while minimizing redundancy. It involves calculating scores for each feature and selecting the top-$k$ features that are most beneficial for the model. Out of the two primary methods explored, namely the Chi-Square method \cite{Thaseen} and Minimum Redundancy Maximum Relevance, we opt for the Chi-Square method. The Chi-Square or $\chi^2$ method (Eq. \ref{chi}) employs a statistical test to assess how features deviate from an expected distribution. In the Eq. \ref{chi}, $O$ and $E$ represents observed and expected frequency/value respectively. $\chi^2$ assigns a variance-based score to each feature, indicating its relevance relative to other features, effectively selecting the best-$k$ features through a threshold mechanism.

\begin{table}[t!]
\renewcommand{\arraystretch}{1.5}
\centering
\caption{Practice Run Results}
\scalebox{0.85}{
\begin{tabular}{|c|c|c|c|c|c|}
    \hline
    \thead{Sr No.} & \thead{Driving Style} & 
    \thead{Method} & \thead{Velocity\\ (\si{\km\per{\hour}})} 
    & \thead{Current \\ (\si{\ampere}) }
    & \thead{Efficiency \\ (\si{\km\per{\kilo\watt\hour}})} \\ 
    \hline
    \multirow{3}{*}{1} & \multirow{3}{*}{Acceleration} & \makecell{Aggressive} & 43.2-10.8 & 40 & 62.5 \\ 
    &  & \makecell{Gentle} & 36.0-12.2 & 32 & 107.1 \\ 
    &  & \makecell{Mild} & 25.0-10.5 & 25 & 78.9 \\ 
    \hline
    \multirow{2}{*}{2} & \multirow{2}{*}{Burn \& Coast} & \makecell{Full} & 32.0-15.3 & 27 & 150.2\\ 
    & & \makecell{Mild} & 23.4-10.8 & 20 & 125.5 \\ [0.5ex]
    \hline
    3 & Pace Feedback & \thead{TEMSL \\ Burn \& Coast} & 28.6-15.1 & 24 & 166.7 \\[0.5ex]
    \hline
\end{tabular}
}
\label{Eval}
\end{table}

\begin{figure*}[htb]
    \centering
    \includegraphics[scale=0.183]{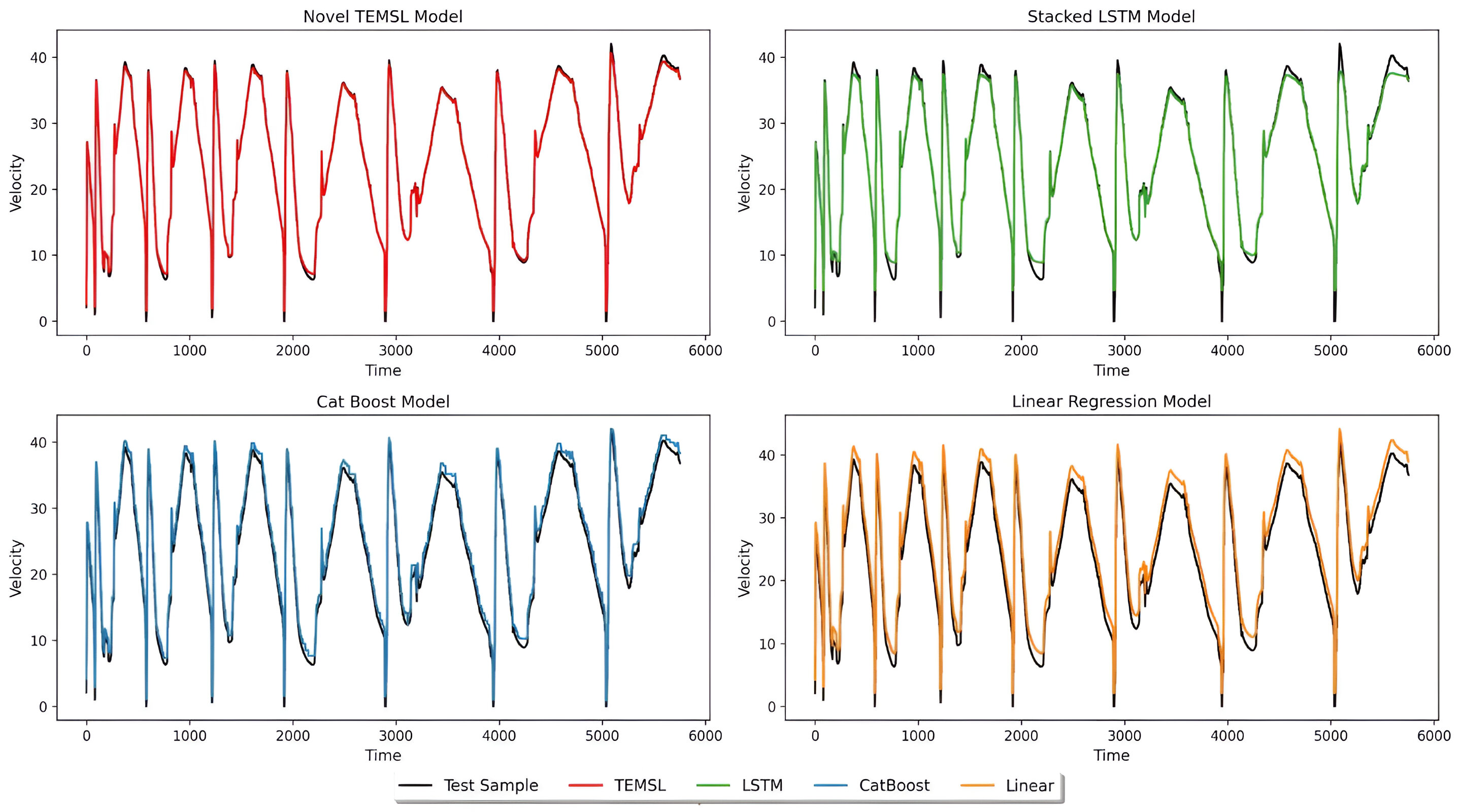}
    \caption{Graphical Representation of the True and Predicted values using the various baseline algorithms}
    \label{Prediction}
\end{figure*}

\subsection{Modelling Framework} 
A deep learning (DL) framework has been applied for our proposed heterogeneous ensemble model \cite{Mangla21,Valentini} to predict the optimal velocity of our vehicle. A heterogeneous ensemble model is a ML approach that combines predictions from multiple diverse base algorithms. This diversity enhances the overall model's performance and robustness by reducing over-fitting. We have implemented our ensemble model (TEMSL), acronym for Tree-based Ensemble Model with stacked LSTM algorithm, using three primary algorithms: Random Forest Regressor, Extra Trees Regressor, and custom Stacked Deep LSTM Network using bootstrap aggregation. The Random Forest excels in handling structured data and complex interactions, Extra Trees Regression is known for incorporating randomization of decision trees to enhance model robustness, and LSTM networks are proficient in modelling sequential data with memory retention. The proposed model is typically compared across three major baseline approaches, as listed below.
\begin{itemize}
    \item {CatBoost Regression: }{CatBoost is a ML supervised algorithm known for combining ordered boosting trees, which prevents overfitting to deliver enhanced results.}
    \item{Stacked Deep LSTM Network: }{This is a novel DL approach that employs multiple LSTM layers which exceeds in capturing sequential dependencies.}
    \item{Linear Regression (LR): }{LR is the most popular supervised regression algorithm that captures the linear relationship between the features and data labels. }
\end{itemize}
The architecture or framework of our proposed model has been illustrated in the Figure \ref{flowchart}. Our proposed method performs well on all performance metrics, surpassing the baseline algorithms utilized in this research study.

\subsection{Model Evaluation}
The experimentation process was done using the TensorFlow Keras deep learning library \cite{Nonsiri} in Python Language. All the algorithms were trained under the exact hyper-parameters as the other baseline methods. The dataset comprises of more than $14000$ entries, which were finally inhibited to a few relevant features using the feature extraction and selection techniques. The split ratio in the dataset is 0.3, which comprised of 70\% data for training and 30\% data for testing. After extensive training process, the experimented algorithms were evaluated fairly on the test samples, as shown in Table \ref{Losses}, using the following evaluation metrics. 
\begin{itemize}
    \item {Mean Absolute Error (MAE): }{Measures the absolute magnitude of errors between predicted and actual values to provide a judgement of model accuracy, with smaller MAE values indicating better performance.}
    \item {Mean Squared Error (MSE): }{Computes the mean of the squared discrepancies between forecasted and real values. It heavily penalizes bigger mistakes, which makes it susceptible to anomalies.}
    \item {R\textsuperscript{2} Score (R\textsuperscript{2}): }{Quantifies how well a model explains the variance in the data. It ranges between 0 and 1, with higher values indicating a better fit. R\textsuperscript{2} helps evaluate the fitness of regression models.}
\end{itemize}

\section{Results} 
Leveraging a dataset collected through rigorous test track practice sessions, we introduced the innovative TEMSL approach, focusing on accurately predicting the optimal vehicle speed to optimize energy efficiency. After meticulous experimentation (Table \ref{Eval}, \ref{Losses}), TEMSL outstripped the existing baseline approaches across various performance metrics. Most notably, TEMSL achieved an exceptional validation accuracy (R\textsuperscript{2}-Score) of 99.85\%, affirming its potential to significantly enhance energy efficiency in real-world vehicular applications. The visual representation of the comparison between models is illustrated in Fig. \ref{Prediction}. The results from the model seemed favourable to our goal of achieving the maximum energy-efficiency possible, ultimately sending feedback to the driver to precisely follow the chosen strategy. \\By installing this model on the Raspberry Pi, which controls the Steering Wheel (Feedback Mechanism) and applying the TEMSL algorithm with the full Burn and Coast driving method, we observed significant improvements in efficiency. After training and testing on the practice run data, the new test data from the final run was fed into the algorithm, and achieved an accuracy of $99.85\%$. The model compared the predicted value with the actual value. When the velocity exceeded the reference speed by 5 \si{\km\per{\hour}}, a green light illuminated, signalling the driver to coast (decelerate) for optimal speed. In the alternate case, the light turned red. If the actual velocity was within the range of $ \pm 2.5 $  \si{\km\per{\hour}} of the predicted value, the Pace Feedback turned blue. The best efficiency achieved by the vehicle was 166.7 \si{\km\per{\kilo\watt\hour}}, which was 26.5 \si{\km\per{\kilo\watt\hour}} more than the total Burn \& Coast efficiency.\\ 

\begin{center}
\begin{table}[t!]
\centering
\caption{Performance Metrics}
\scalebox{0.95}{
\begin{tabular}{|c|c|c|c|}
    \hline
    \thead{Algorithms} & \thead{MAE} & \thead{MSE} & \thead{R\textsuperscript{2} Score} \\ 
    \hline
    \makecell{TEMSL} & 0.17 &  0.08 & 99.52\% \\ 
    \hline
    \makecell{Stacked LSTM} & 0.46 &  0.56 & 98.15\% \\ 
    \hline
    \makecell{CatBoost} & 0.71 &  0.92 & 96.65\% \\ 
    \hline
    \makecell{Linear Regression} & 3.12 &  14.18 & 86.45\%\\ 
    \hline  
\end{tabular}
}
\label{Losses}
\end{table}
\end{center}

\vspace{-32pt}

\section{Conclusion}
Through the process of data acquisition, exploratory data analysis, and computation, the study of the effect of driving techniques on the overall vehicle efficiency was carried out successfully. The onboard real-time telemetry system is actively assisting the driver in adjusting driving patterns for minimum energy consumption. The novel approach, TEMSL, assisted us in accurately selecting the suitable driving approach based on sensor and haptic feedback during track driving. Using the suggested Pace feedback mechanism equipped with TEMSL approach, an increase in energy efficiency of about 10.6\% was recorded as compared to the closest alternative technique, thus validating our hypothesis. The current research focuses and analyzes primarily two driving patterns commonly used for minimizing energy consumption. However, since our research is centered on just two driving patterns, further research work needs to be conducted to impact the energy economy by the use of intelligent systems for various types of driving behaviors. The model also has the potential to contribute to efficiency enhancements on real city roads, beyond our chosen test track. This research can foster a broader societal transition towards cleaner and greener transportation solutions by demonstrating the successful implementation of the mechanism in enhancing the performance and energy efficiency of EVs.

\section*{Acknowledgment}
This work was supported and funded by Delhi Technological University (DTU), New Delhi, India. Furthermore, we extend our gratitude to Shell Eco-Marathon Committee for providing us an international platform to showcase our research innovations.

\vspace{12pt}


\begin{thebibliography}{00}
\bibitem{b1} Masoumi, Houshmand E., and Ahmadreza Faghih Imani. "Energy Consumption and Transportation in Developing Countries: Need for Local Scenario-Based Energy Efficiency Plans." Sustainability 7, no. 3 (2015): 2560-2583
\bibitem{b2} B. Bolaji and S. B. Adejuyigbe, "Vehicle Emissions and their Effects on the Natural Environment - a Review," Journal of the Ghana Institution of Engineers, vol. 4, pp. 35-41, 2006.
\bibitem{b3} Canan G. Corlu, Rocio de la Torre, Adrian Serrano-Hernandez, Angel A. Juan, Javier Faulin, ``Optimizing Energy Consumption in Transportation: Literature Review, Insights, and Research Opportunities``, Energies (Multidisciplinary Digital Publishing Institute)-Vol. 13 (2020)
\bibitem{Parekh} D. Parekh et al., “A Review on Autonomous Vehicles: Progress, Methods and Challenges,” Electronics, vol. 11, no. 14, p. 2162, Jul. 2022
\bibitem{Vasudevan} K. Vasudevan, A. P. Das, Sandhya B and Subith P, "Driver drowsiness monitoring by learning vehicle telemetry data," 2017 10th International Conference on Human System Interactions (HSI), Ulsan, Korea (South), 2017, pp. 270-276.
\bibitem{b4} Tobias Bär, Ralf Kohlhaas, J. Marius Zöllner and Kay-Ulrich Scholl, "Anticipatory driving assistance for energy efficient driving," Integrated and Sustainable Transportation System (FISTS), 2011 IEEE Forum , June 2011.
\bibitem{b5} Tianyi Guan and Christian W. Frey, ``Fuel efficiency driver assistance system for manufacturer independent solutions,'' 2012 15th International IEEE Conference on Intelligent Transportation Systems, September 2012.
\bibitem{Triwiyatno} Aris Triwiyatno, Suroto Munahar, Munadi Muna and Joga Dharma Setiawan, ``Application of driving behavior control system using artificial neural network to improve driving comfort by adjusting air-to-fuel ratio", IIUM Engineering Journal 24(2), pp. 337--353, July 2023.
\bibitem{b8} Tianyi Guan and Christian W. Frey, ``Model adaptive driver assistance system to increase fuel savings,'' Vehicular Electronics and Safety (ICVES), 2012 IEEE International Conference, July 2012.
\bibitem{b12} Sandareka Wickramanayake and H.M.N. Dilum Bandara, ``Fuel consumption prediction of fleet vehicles using Machine Learning: A comparative study,'' 2016 Moratuwa Engineering Research Conference (MERCon), April 2016.
\bibitem{Ivarsson} Erik Hellström, Maria Ivarsson, Jan Äslund and Lars Nielsen, ``Look-ahead control for heavy trucks to minimize trip time and fuel consumption,''  Advances in Automotive Control, IFAC Proceedings Volumes 40(10), pp. 439--446, August 2007.
\bibitem{Santana} Jose del C. Julio-Rodriguez, Alfredo Santana, Carlos A. Rojas-Ruiz, Rogelio Bustamante-Bello and Ricardo A. Ramirez-Mendoza, ``Environment Classification Using Machine Learning Methods for Eco-Driving Strategies in Intelligent Vehicles,'' Applied Sciences 12(11):5578, May 2022.
\bibitem{Meseguer} Javier E. Meseguer, Carlos T. Calafate, Juan Carlos Cano and Pietro Manzoni, ``Characterizing the Driving Style Behavior using Artificial Intelligence Techniques,'' 38th IEEE Conference on Local Computer Networks (LCN 2013), October 2013.
\bibitem{Mosabbir} Mosabbir Bhuiyan and Md Ahasan Kabir, ``Vehicle Speed Prediction based on Road Status using Machine Learning,'' Advanced Science, Engineering and Medicine 2(1), pp. 1--9, January 2020.
\bibitem{Pan22} Y. Li and Y. Pan, "A novel ensemble deep learning model for stock prediction based on stock prices and news," Int J Data Sci Anal, vol. 13, pp. 139–149, 2022.
\bibitem{Ravikumar20} S. Ravikumar and P. Saraf, "Prediction of Stock Prices using Machine Learning (Regression, Classification) Algorithms," pp. 1-5, 2020.
\bibitem{Kohlhaas11} Ralf Kohlhaas, Thomas Schamm, Dennis Nienhüser and J. Marius Zöllner, ``Anticipatory energy saving assistant for approaching slower vehicles,'' 2011 14th International IEEE Conference on Intelligent Transportation Systems (ITSC), October 2011.
\bibitem{Wickramanayake} Sandareka Wickramanayake and H.M.N. Dilum Bandara, ``Fuel consumption prediction of fleet vehicles using Machine Learning: A comparative study,'' 2016 Moratuwa Engineering Research Conference (MERCon), April 2016.
\bibitem{Rayamajhi16} K. C. Dey, A. Rayamajhi, M. Chowdhury, P. Bhavsar, and J. Martin, "Vehicle-to-vehicle (V2V) and vehicle-to-infrastructure (V2I) communication in a heterogeneous wireless network – Performance evaluation," Transportation Research Part C: Emerging Technologies, vol. 68, pp. 168-184, 2016. ISSN: 0968-090X.
\bibitem{Younes22} M. B. Younes, "Towards Green Driving: A Review of Efficient Driving Techniques," World Electric Vehicle Journal, vol. 13, no. 6, p. 103, Jun. 2022.
\bibitem{Jeongwoo} Jeongwoo Lee , "Vehicle Inertia Impact on Fuel Consumption of
Conventional and Hybrid Electric Vehicles Using Acceleration and Coast Driving Strategy", Virginia Tech, March 2014.
\bibitem{b14} Syed shujat Ali, ``MPU-6050 (Accelerometer and Gyroscopic sensor),'' Phil. Trans. Roy. Soc. London, vol. A247, pp. 529--551, July 2021.
\bibitem{b15} Marco Crescentini, Sana Fatima Syeda and Gian Piero Gibiino, ``Hall-Effect Current Sensors: Principles of Operation and Implementation Techniques,''  IEEE Sensors Journal, Vol. 22, Iss. 11, pp. 10137--10151, June 2022.
\bibitem{b16} Syed shujat Ali, ``NEO-6M-0-001(GPS module),'' National University of Sciences and Technology, Formula Student Team, July 2021.
\bibitem{b17} Li Tianze, Zhang Xia, Jiang Chuan and Hou Luan, "Analysis of the characteristics of piezoelectric sensor and research of its application," 2009 18th IEEE International Symposium on the Applications of Ferroelectrics, pp. 1-4, August 2009
\bibitem{b18} DTutunea Dragos, Ilie Dumitru, George Gherghina and Alexandru Dima, ``Evaluation of Thermistors Used for Temperature Measurement in Automotive Industry,'' Applied Mechanics and Materials, vol. 880, pp. 157--162, March 2018.
\bibitem{Liao12} S. H. Liao, P. H. Chu, and P. Y. Hsiao, "Review: Data mining techniques and applications - A decade review from 2000 to 2011," Expert Systems with Applications: An International Journal, vol. 39, pp. 11303-11311, 2012. 
\bibitem{Griffiths} P. Taylor, N. Griffiths, A. Bhalerao, S. Anand, T. Popham, Z. Xu, and A. Gelencser, "Data Mining for Vehicle Telemetry," Applied Artificial Intelligence, vol. 30, no. 3, pp. 233-256, 2016.
\bibitem{Constantinescu} Z. Constantinescu, C. Marinoiu, and M. Vladoiu, "Driving style analysis using data mining techniques," International Journal of Computers Communications \& Control, vol. 5, no. 5, pp. 654–663, 2010.
\bibitem{Adekoya} I. K. Nti, A. F. Adekoya, and B. A. Weyori, "A comprehensive evaluation of ensemble learning for stock-market prediction," J Big Data, vol. 7, p. 20, 2020.
\bibitem{Rosipal} R. Rosipal, M. Girolami, L. Trejo, et al., "Kernel PCA for Feature Extraction and De-Noising in Nonlinear Regression," NCA 10, pp. 231–243, 2001.
\bibitem{Thaseen} S. Thaseen and A. K. Cherukuri, "Intrusion Detection Model Using fusion of Chi-square feature selection and multi class SVM," Journal of King Saud University - Computer and Information Sciences, vol. 29, 2016.
\bibitem{Mangla21} N. Sharma, J. Dev, M. Mangla, V. Wadhwa, S. Mohanty, and D. Kakkar, "A Heterogeneous Ensemble Forecasting Model for Disease Prediction," New Generation Computing, vol. 39, 2021.
\bibitem{Valentini} M. Re and G. Valentini, "Ensemble methods: A review," 2012.
\bibitem{Nonsiri} J. F. J. Joseph, S. Nonsiri, and A. Monsakul, "Correction to: Keras and TensorFlow: A Hands-On Experience," in Advanced Deep Learning for Engineers and Scientists, edited by K. B. Prakash, R. Kannan, S. Alexander, G. R. Kanagachidambaresan, EAI/Springer Innovations in Communication and Computing, Springer, Cham, 2021.





\end{thebibliography}
\end{document}